\def\Murcia{Departamento de Matem\'atica Aplicada, Facultad de Inform\'atica, Campus
de Espinardo, 30100 Murcia, Spain}

\def\IAA{Instituto de Astrofísica de Andalucía (CSIC), Apartado Postal 3004,
18080 Granada, Spain}

\def\CarlosI{Instituto de F\'\i sica Te\'orica y Computacional Carlos I,
Facultad de Ciencias, Universidad de Granada, Campus de Fuentenueva,
Granada 18002, Spain}
\def\Comision{Work partially supported by the DGICYT.}

\def\bdm{\begin{displaymath}}
\def\edm{\end{displaymath}}
\def\bea{\begin{eqnarray}}
\def\eea{\end{eqnarray}}
\def\be{\begin{equation}}
\def\ee{\end{equation}}
\def\ba{\begin{array}}
\def\ea{\end{array}}

\def\noi{\noindent}
\def\nn{\nonumber}

\def\medio{\frac{1}{2}}

\def\HNln{H_n^N}

\newcommand{\parcial}[1]{ \frac{\partial}{\partial #1} }

\documentclass[12pt]{article}
\usepackage{amsfonts,amssymb}

\begin{document}


\begin{center}
{\LARGE {\bf Group Approach to the Quantization of the P\"oschl-Teller dynamics$^1$ }}
\end{center}

\vskip 0.2 cm

\centerline{ V. Aldaya$^{2,3}$ and J. Guerrero$^{2,3,4}$}

\footnotetext[1]{\Comision} \footnotetext[2]{\IAA}
\footnotetext[3]{\CarlosI} \footnotetext[4]{\Murcia}

 \begin{center}
{\bf Abstract}
\end{center}
\small
 \begin{list}{}{\setlength{\leftmargin}{3pc}\setlength{\rightmargin}{3pc}}
\item The quantum dynamics of a particle in the Modified P\"oschl-Teller potential is derived
from the group $SL(2,\mathbb{R})$ by applying a Group Approach to Quantization (GAQ).
The explicit form of the Hamiltonian as well as the ladder operators is found in the enveloping algebra 
of this basic symmetry group. The present algorithm provides a physical realization of
the non-unitary, finite-dimensional, irreducible representations of the $SL(2,\mathbb{R})$ group.
The non-unitarity manifests itself in that only half of the states are normalizable, in contrast with
the representations of $SU(2)$ where all the states are physical.
\end{list}
\normalsize

\section{Introduction}
Symmetry has proven very useful in Quantum Mechanics as a powerful tool to construct
explicitly the eigenstates and eigenvalues of a given symmetrical Hamiltonian. Since
the pioneering work of Wigner \cite{Wigner} many papers have been devoted to the analysis
of solvable quantum systems through their ``dynamical symmetries'' or
``spectrum-generating algebras'' \cite{B-N-B}. In particular, the P\"oschl-Teller and
Morse potentials, bounding molecular systems, have been soundly studied along these
lines \cite{Poschl-Teller,Rosen-Morse,Iachello,Wolf} (see also \cite{Arias} for recent and more
detailed bibliography). But symmetry can be taken beyond this ability and constitutes
the fundamentals for physical systems in such a way that any referent to them, that is,
space-time, classical solution manifold, wave functions, operators, scalar product, etc,
can be explicitly derived in a natural manner from a particular Lie group. This
viewpoint has been demonstrated in many finite- and infinite-dimensional cases by
applying a Group Approach to Quantization developed since the original paper \cite{23},
where the quantum free Galilean particle and the harmonic oscillator were derived. Then,
this algorithm has been applied to less elementary groups as those associated with
relativistic particles, in particular the relativistic harmonic oscillator \cite{oscilata,RMP,oscipert},
field theories in curved space-times, non-linear $\sigma$-models, the Virasoro group and
others concerning conformal symmetry and quantum gravity (see, for instance
\cite{frachall,medida,vacuum}).

The Modified P\"oschl-Teller potential (MPT), however, has a special attractive in spite of its simplicity,
because it seems not to be primarily associated with a particular symmetry but, rather, with a
phenomenological force, and it is less integrable than other more involved physical problems.
In specific terms, the classical Hamiltonian does not close a Poisson subalgebra with the
coordinate and the momentum. This system prompts us to search for an alternative
finite-dimensional Poisson subalgebra in the free algebra generated by $\langle H,x,p\rangle$, a procedure
which would be of a wide usage since these generators generally fail in closing a subalgebra in
many physical systems. In fact, it is possible to find two classical functions,
${\cal X}$ and ${\cal P}$, that close with the classical Hamiltonian $H$ a ``Lie algebra'' having
the structure constants
replaced with functions of the energy, a breakdown of the Lie structure similar to that found in
the Hydrogen atom when trying to close the dynamical symmetry  generated by
the angular momentum and the Runge-Lenz vector \cite{Shiff,Bacry}, which  turns out to be
$SO(4)$, $E(3)$ or $SO(3,1)$ depending on whether the fixed energy is negative, null or positive. 
Unlike the Hydrogen atom, here the energy is not central in these Lie algebras, and
the symmetry proves to be $SO(2,1)$ (or $SU(1,1)\approx SL(2,\mathbb{R})$) although 
for bounded states (negative energy) a complex prolongation of this algebra can be confused
with $SU(2)$ at the classical level. In fact, since the Lie algebras of these groups have the same 
complex form, a complex prolongation from one part of the spectrum to the other can be easily performed.

We shall proceed by taking the square root of $H$ and considering the set
$\langle{\cal E}\approx\sqrt{H},{\cal X},{\cal P}\rangle$, which close a true Lie algebra, $SO(2,1)$, as the
starting point for the GAQ. From an algebraic group law for $SO(2,1)$ we derive the unitary
irreducible representations of the group as well as the explicit expression of
all operators in the (enveloping) algebra  and, in particular, the operator ${\cal E}^2\approx H$. This
operator will results in $\hat{p}^2-D/\hbox{cosh}^2(\alpha x)$, i.e. the quantum operator representing the
original Hamiltonian associated with the potential $V(x)=-D/\hbox{cosh}^2(\alpha x)$, $D$ being the
absolute value of the potential depth and $\alpha$ an indicative of its width.

A remarkable feature appearing in the present process is that the eigenstates of the
MPT Hamiltonian with negative energy, i.e. the bound states, are formally
obtained from the wave functions in the discrete series of the $SL(2,\mathbb{R})$ unitary
irreducible representations (the wave functions for a model of a relativistic harmonic oscillator)
with negative Bargmann index $-q<0$. This can be seen to correspond \cite{Lang} to a non-unitary, 
finite-dimensional, representation of $SL(2,\mathbb{R})$ corresponding to positive Bargmann index $q$.
The non-unitarity of the representation reveals itself in the fact that not all the wave-functions
are normalizable, and therefore the physical Hilbert space is smaller. In fact, from the $2q+1$ states
of the representation, $\Psi^q_n,\,,n=0,\ldots,2q$, only $[q]+1$ are normalizable 
(where $[q]$ stands for the smaller, closest integer to $q$). If $q$ is an integer,
there are only $q$ states, from $n=0,\ldots,q-1$, the state with $n=q$ (which correspond to zero energy)
being not normalizable. If $q$ is half integer, there are $q+\medio$ states, from $n=0,\ldots,q-\medio$.
Going to the universal covering group of $SL(2,\mathbb{R})$ real values of $q$ are also allowed.

This behavior is very different form that of $SU(2)$, where the representations are $2j+1$ dimensional,
 with $j$ integer or half-integer,
but all states are normalizable since the representations are unitary. This shows that the correct
symmetry for bounded states is not $SU(2)$, as it is normally claimed in the literature
(see, for instance, \cite{Iachello,Arias}), but, rather the finite-dimensional representations of
$SL(2,\mathbb{R})$. Note that the quantum description of the MPT system in terms of $SU(2)$ real values
of $q$ would be forbidden.

This paper is organized as follows. In Sec. 2 the classical dynamics in the MPT
potential is presented aiming at finding the relevant symmetry that will be quantized in
abstract terms in the framework of GAQ. Sec. 3 is devoted to a very brief report on GAQ and,
finally, the quantum dynamics associated with the MPT interaction, as well as the corresponding
Hamiltonian and ladder operators, is derived in Sec. 4.

\section{Classical theory and Poisson symmetry}
\label{}
Even though GAQ is primarily intended to achieve quantum systems without the previous step
of solving the classical counterpart, the classical theory can help us in finding the relevant
symmetry. Then, we proceed to solve the classical equations of motion and to look for an
appropriate symmetry as an input to GAQ.

The Lagrangian for the MPT potential, with positive depth $D$ and width $1/\alpha$,
can be written as
\be
L=\frac{1}{2}m\dot{x}^2+\frac{D}{\hbox{cosh}^2(\alpha x)}=\frac{1}{2}m\frac{\dot{\xi}^2}
{1+\alpha^2\xi^2}+\frac{D}{1+\alpha^2\xi^2}\,,
\ee
\noi where we have introduced the coordinate $\xi=\frac{\hbox{sinh}(\alpha x)}{\alpha}$.

Let us solve the Euler-Lagrange equations for negative energy $E=-\epsilon$, $\epsilon>0$. They are:
\be
\dot{\xi}=\sqrt{\frac{2}{m}\left[(1+\alpha^2\xi^2)E+D\right]}
\ee
\noi i.e.
\be
\frac{d\xi}{\sqrt{\frac{2\epsilon}{m}}\sqrt{\left(\frac{D-\epsilon}{\epsilon}\right)-\alpha^2\xi^2}}=dt\;,
\ee
\noi the solution to which is:
\be
\xi=\sqrt{\frac{D-\epsilon}{\alpha^2\epsilon}} \hbox{sin}\left(\sqrt{\frac{2\epsilon\alpha^2}
{m}}t+\phi_0\right)\;,
\ee
\noi where $\phi_0\equiv\hbox{sin}^{-1}\frac{\alpha\xi_0}{\sqrt{\frac{D-\epsilon}{\epsilon}}}$
is the initial phase. Writing also the equation for the velocity we arrive at a couple of equations,
\bea
\xi&=&\xi_0\hbox{cos}\sqrt{\frac{2\epsilon\alpha^2}{m}}t+
\sqrt{\frac{m}{2\epsilon\alpha^2}}\,\dot{\xi}_0\,\hbox{sin}\sqrt{\frac{2\epsilon\alpha^2}
{m}}t \nn\\
\dot{\xi}&=&\dot{\xi}_0\,\hbox{cos}\sqrt{\frac{2\epsilon\alpha^2}{m}}t
-\sqrt{\frac{2\epsilon\alpha^2}{m}}\,\xi_0\,\hbox{sin}\sqrt{\frac{2\epsilon\alpha^2}{m}}t
 \;,\label{eqPT}
\eea
\noi where $\dot{\xi}_0\equiv\sqrt{\frac{2\epsilon\alpha^2}{m}}\sqrt{\frac{D-\epsilon}{\alpha^2\epsilon}
-\xi^2_0}$ is the initial velocity. They go to those of the harmonic oscillator in the limit
in which $D\rightarrow\infty$, $\alpha\rightarrow 0$, but $\frac{2D\alpha^2}{m}$ is kept finite and
equal to $\omega^2$
(constant), that is:
\bea
\xi&=&\xi_0\hbox{cos}\omega t+\frac{\dot{\xi}_0}{\omega}\hbox{sin}\omega t\nn\\
\dot{\xi}_0&=&\dot{\xi}_0\hbox{cos}\omega t-\omega\xi_0\hbox{sin}\omega t\;.
\eea

Equations (\ref{eqPT}) behave as those of an harmonic oscillator with a frequency depending on
the energy. In fact, the Hamiltonian can be written as:
\bea
H &=& \frac{1}{2}m\frac{\dot{\xi}^2}{1+\alpha^2\xi^2}-
\frac{D}{1+\alpha^2\xi^2} =
\frac{1}{2}m\frac{\dot{\xi}^2}{1+\alpha^2\xi^2}+ \frac{D\alpha^2\xi^2}{1+\alpha^
2\xi^2} -D \nn \\
& = & \frac{1}{2}m \dot{\xi}^2+\frac{1}{2}m \omega(\epsilon)^2\,\xi^2 -D
\eea
\noi where $\omega(\epsilon)\equiv \sqrt{\frac{2\epsilon\alpha^2}{m}}$, and this,
up to the constant energy shift $-D$, is an harmonic oscillator with energy-dependent frequency
$\omega(\epsilon)$\footnote{See \cite{delOlmo} for a unified derivation of different integrable potentials in one and two dimensions and a description
of their solutions, among them the different versions of the P\"oschl-Teller potentials. See also
\cite{Carinena} for an interpretation of this system as a harmonic oscillator with position dependent mass.}. 
For positive energy they transform into the equations of motion for a
``repulsive-like'' oscillator.

In order to write the Poisson bracket we observe the Poincar\'e-Cartan form $\Theta_{PC}=
pdx-Hdt=\frac{\partial L}{\partial\dot{x}}(dx-\dot{x}dt)+Ldt$:
\bea
\Theta_{PC}&=&\frac{m\dot{\xi}}{1+\alpha^2\xi^2}d\xi-\left(\frac{1}{2}m\frac{\dot{\xi}^2}{1+\alpha^2\xi^2}-
\frac{D}{1+\alpha^2\xi^2}\right)dt\nn\\
&=&p_\xi d\xi-\left[(1+\alpha^2\xi^2)\frac{p_\xi^2}{2m}-\frac{D}{1+\alpha^2\xi^2}\right]dt\;,\label{PC}
\eea
\noi where the momentum canonically conjugate to $\xi$ is
\be
p_\xi\equiv\frac{\!\partial L}{\partial\dot{\xi}}=\frac{m\dot{\xi}}{1+\alpha^2\xi^2}\;.
\ee
\noi A simple inspection of $\Theta_{PC}$ indicates that the basic Poisson bracket will acquire the
canonical form:
\be
\{\xi,\,p_\xi\}=1\;.
\ee

By examining the Poisson bracket of $H$ with $\xi$ and $p_\xi$ we observe that
$\{H,\xi,(1+\alpha^2\xi^2)p_\xi\}$ ``close'' a Lie subalgebra with structure constants
depending on $H$, and that it is possible to close a true algebra by choosing an
appropriate function of $H$ to replace $H$. To be precise, the following
classical functions close a $SO(2,1)$ algebra:
\be
<{\cal E}\equiv 2\sqrt{D}\sqrt{H},\;{\cal X}\equiv\frac{\sqrt{2}}{\sqrt{D}}\sqrt{H}\xi,\;
{\cal P}\equiv\sqrt{2}(1+\alpha^2\xi^2)p_\xi>
\ee
\noi In fact, we find:
\bea
\{{\cal E},{\cal P}\}&=&-m\Omega^2{\cal X}\nn\\
\{{\cal E},{\cal X}\}&=&-\frac{1}{m}{\cal P} \label{SO21} \\
\{{\cal X},{\cal P}\}&=&\frac{1}{D} {\cal E}\,,\nn
\eea
\noi where $\Omega\equiv\sqrt{\frac{2\alpha^2D}{m}}=\omega(D)$, which is the frequence of the small 
oscilations (harmonic approximation).

For positive energy (scattering states) ${\cal E}$ can be diagonalized in terms of real
combinations of ${\cal X}$ and ${\cal P}$, $\langle A \equiv \frac{1}{\Omega}{\cal E},
\;B \equiv \frac{1}{2\alpha}\left( {\cal P}+m\Omega {\cal X}\right),
\;C \equiv \frac{1}{2\alpha}\left( {\cal P}-m\Omega {\cal X}\right)\rangle,\;$
giving rise to the standard form of the $SL(2,\mathbb{R})$ algebra:
\bea
\{A,B\}&=&-B\nn\\
\{A,C\}&=&C \label{SL2R} \\
\{B,C\}&=& A\,.\nn
\eea

However, we are interested in describing bounded states, with negative energy. For this states,
${\cal E}$ and ${\cal X}$ are
pure imaginary and, therefore, we must redefine ${\cal E}'\equiv -i{\cal E},\; {\cal X}'\equiv -i{\cal X}\;$ and
$\;{\cal P'}\equiv {\cal P}$. Then, the complex combinations
$<L_0 \equiv \frac{1}{\Omega}{\cal E}',
\;L_- \equiv \frac{1}{2\alpha}\left( {\cal P}'-im\Omega {\cal X}'\right),
\;L_+\equiv \frac{1}{2\alpha}\left( {\cal P}'+im\Omega {\cal X}'\right)>$
satisfy the algebra:
\bea
\{L_0,L_+\}&=&iL_+\nn\\
\{L_0,L_-\}&=&-iL_- \label{SU2}\\
\{L_+,L_-\}&=&i L_0\nn
\eea
\noi which can be identified with the complex form of both $SU(2)$ and $SU(1,1)$ algebras.
Observing carefully the different algebras, it can be realized that $L_+=B$, $L_-=C$ and $L_0=-iA$. This
means that the two diagonalizations are the same, the only difference being the use of $A$ or $L_0$, which
are real for positive and negative energies, respectively. This allows us to consider the two cases
simultaneously, with a single algebra, $SO(2,1)$ or its different versions $SL(2,\mathbb{R})$ or $SU(1,1)$,
for both positive and negative energies, instead of using the $SU(2)$ algebra for describing bounded
states. The only difference will
lie in the fact that for negative energies, some generators will be non-hermitian,
or the pair of creation-annihilation operators will not be the adjoint to each other, in other words, the
representation obtained will fail to be unitary. We shall postpone the discussion of its implications
to the final section.



\section{Group Approach to Quantization}
\subsection{Brief report on the general theory}
The group approach to quantization \cite{23,Ramirez,medida,Marmo} lies on the simple idea that
the essential of a quantum theory is nothing other than a unitary irreducible representation of a
Lie algebra usually, though non-necessarily, associated with a Poisson subalgebra of the
solution manifold of a classical system. The GAQ algorithm constitutes simply a technique for
representing Lie groups in a geometric way using only canonical structures on a Lie group,
the quantum states being complex functions on the group manifold itself. The carrier
space supports the realization of all operators (and only those) in the enveloping algebra.

Let us remind the reader that on any Lie group with composition law
$g''=g'*g$, two different and compatible actions can be
considered. In fact, the left and right actions
\bea
L_{g'}: G&\rightarrow& G\nn\\
g&\mapsto& g''\nn\\
R_g:G&\rightarrow& G\nn\\
g'&\mapsto&g''\nn
\eea
\noi are generated by right-invariant and left-invariant vector fields on
the group $G$, ${\cal X}^R$ and ${\cal X}^L$ respectively, and both commute. This is a
remarkable property which allows us to adopt one of those Lie algebras,
let us say ${\cal X}^R$, as well as the associated enveloping algebra,
as the set of physical operators $\hat{g}^i\sim X^R_{g^i}$, whereas the other is used to reduce the
corresponding representation in a compatible way, by nullifying a maximal
subalgebra (in the left enveloping algebra, in general),  named {\it polarization}, on the (reduced)
wave functions. Mostly, the relevant symmetry group is a central extension, $\tilde{G}$ of a Lie group
$G$ by $U(1)$. Aiming at representing the canonical Poisson bracket between $x$ and $p$, the complex
functions on $\tilde{G}$ are then prompted to satisfy the $U(1)$-constraint
\be
\tilde{X}^R_\phi\Psi=i\Psi\;.\label{U1}
\ee

The classical theory, including the space of coordinates, momenta and time, is recuperated out of
the group manifold in a manner similar to the way we obtain the solution manifold from
the $(q,p,t)$ space as the quotient by the kernel of the differential of the Poincar\'e-Cartan form
$\Theta_{PC}$. In fact, there is a generalized Poincar\'e-Cartan form on
the group, the quantization 1-form $\Theta$, such that
${\cal P}\equiv \tilde{G}/(\hbox{Ker}d\Theta\cap\hbox{Ker}\Theta)$ is a quantum manifold in the sense of
Geometric Quantization \cite{GQ}, and ${\cal S}\equiv \tilde{G}/\hbox{Ker}d\Theta$ is
the classical solution manifold. ${\cal S}$ can be parameterized by
functions of the form $\Theta(X^R_{g^i})$, which are the Noether
invariants.

\subsection{The example of the relativistic harmonic oscillator}
We resort to a rather non-trivial $1+1$-dimensional example to achieve two tasks.
On the one hand we exemplify the GAQ algorithm on a physical system, that is, a
relativistic harmonic oscillator (RHO) or a particle moving on $1+1$-Anti-de Sitter
space-time and, on the other, we arrive at precise results
on the representations of $SU(1,1)\approx SL(2,\mathbb{R})$ that will be required in the next section.
Simpler examples can be found in Ref. \cite{23}.

Quantum symmetry differs from the classical counterpart in an extra phase (or $U(1)$)
transformation which permits the realization of an exact invariance of action
integrands (Lagrangians or Poincar\'e-Cartan forms), versus the semi-invariance
achieved in Classical Mechanics. This is so even in the case of finite-dimensional
semi-simple groups for which all central extensions are mathematically
trivial. In fact, the actual central extension of the Lie algebra of such a symmetry
points out to a specific coadjoint orbit of the classical symmetry and, then, the phase
space (solution manifold) of the classical system \cite{sympl}. Let us comment very
briefly on these details in relation to the case of the free 1+1D non-relativistic and
relativistic particle. The quantum symmetry of the Galilean particle
obeys the following commutation relations (representing the classical Poisson brackets):
\bea
\left[\tilde{X}^R_t,\;\tilde{X}^R_x\right]&=&0\nn\\
\left[\tilde{X}^R_t,\;\tilde{X}^R_p\right]&=&-\frac{1}{m}\tilde{X}^R_x
\label{Galileo}\\
\left[\tilde{X}^R_x,\;\tilde{X}^R_p\right]&=&\tilde{X}^R_\phi\;,\nn
\eea
\noi where $\tilde{X}^R_\phi$ is the central generator associated with the
phase invariance of wave functions, which are constrained to the
$U(1)$-function condition (\ref{U1}) in order to
represent the classical Poisson algebra among $x$, $p$, $\frac{p^2}{2m}$ and $1$. The algebra
(\ref{Galileo}) constitutes a non-trivial central extension of that of
the Galilei group by $U(1)$. In going to the relativistic case, the
Poincar\'e group is also centrally extended, though trivially, in a way
that the corresponding algebra reads:
\bea
\left[\tilde{X}^R_t,\;\tilde{X}^R_x\right]&=&0\nn\\
\left[\tilde{X}^R_t,\;\tilde{X}^R_p\right]&=&-\frac{1}{m}\tilde{X}^R_x
\label{Poincare}\\
\left[\tilde{X}^R_x,\;\tilde{X}^R_p\right]&=&\frac{1}{mc^2}\tilde{X}^R_t+\tilde{X}^R_\phi\;,\nn
\eea
\noi In the non-relativistic limit, this algebra contracts to the
non-trivial extension (\ref{Galileo}).

To describe the quantum $SL(2,\mathbb{R})$ symmetry, physically realized as a
quantum relativistic harmonic oscillator \cite{oscilata,RMP,oscipert} we can dilate (as the
opposite to contract) the algebra (\ref{Poincare}) with an extra term in the
r.h.s. so that it contract to the Poincar\'e algebra in the limit
$\omega\rightarrow 0$ and to the non-relativistic harmonic oscillator,
with angular frequency $\omega$, in the $c\rightarrow\infty$ limit. We shall apply
the group quantization mechanism to the resulting group parameterized with renamed time,
position and momentum variables,  $(\tau,\,y,\pi)$, as well as the mass, $\mu$, to
prevent any confusion with analogous variables in the physical problem
analyzed in the next section. We then write:
\bea
\left[\tilde{X}^R_\tau,\;\tilde{X}^R_y\right]&=&-\mu\omega^2\tilde{X}^R_\pi \nn\\
\left[\tilde{X}^R_\tau,\;\tilde{X}^R_\pi\right]&=&-\frac{1}{\mu}\tilde{X}^R_y
\label{Oscilata}\\
\left[\tilde{X}^R_y,\;\tilde{X}^R_\pi\right]&=&\frac{1}{\mu c^2}\tilde{X}^R_\tau+\tilde{X}^R_\phi\;,\nn
\eea

These commutation relations can be exponentiated to a group law in many
(equivalent) ways, the next one being a possibility:
\bea
 \sin\omega\tau''&=&\frac{\omega}{\beta''}(\frac{\beta}{\mu c^{2}\beta'}\pi'y'
     \sin\omega\tau'\sin\omega\tau+\frac{\beta \Pi_{0}'}{\mu\omega\beta'c}\cos\omega\tau'\sin\omega\tau \nn \\
 & &+\frac{\omega}{\beta'\mu c^{3}}yy'\Pi_{0}'\sin\omega\tau'+\frac{\beta'\beta}{\omega}
     \cos\omega\tau\sin\omega\tau'+\frac{\pi'y}{\mu c^{2}\beta'}\cos\omega\tau') \nn \\
 y''&=&\frac{\pi'\beta}{\mu\omega}\sin\omega\tau+\beta y'\cos\omega\tau+\frac{y\Pi_{0}'}{\mu c}
 \label{RGL1} \\
 \pi''&=&\frac{\omega y\pi}{\beta c^{2}}(\frac{\pi'}{\mu}\sin\omega\tau+\omega y'\cos\omega\tau)
     +\frac{\Pi_{0}}{c\beta}(\frac{\pi'}{\mu}\cos\omega\tau-\omega y'\sin\omega\tau)+\frac{\pi\Pi_{0}'}{\mu c} \nn \\
 \zeta''&=&\zeta'\zeta e^{\frac{i}{\hbar}(\delta''-\delta'-\delta)} \, , \nn
\eea
 \noi where
\bea
 \Pi_{0}&\equiv&\sqrt{\mu c^{2}+\pi^{2}+\mu^{2}\omega^{2}y^{2}} \nn \\
 \beta&\equiv&\sqrt{1+\frac{\omega^{2}y^{2}}{c^{2}}} \nn \\
 \delta&\equiv&-\mu c^{2}\tau-f \label{RGL2}   \\
 f&\equiv&-\frac{2\mu c^{2}}{\omega}\tan^{-1}\left[\frac{\mu c^{2}}
    {\omega\pi y}(\beta-1)(\frac{\Pi_{0}}{\mu c}-\beta)\right] \, . \nn
\eea
\noi From the group law above we derive directly the set of left-invariant
vector fields, which are relevant in the reduction procedure, through a
polarization algebra, and the generalization of the Poincar\'e-Cartan
form,
\bea
 \tilde{X}^{L}_{\tau}&=&\frac{\pi}{\mu}\frac{\partial}{\partial
 y}-\mu\omega^{2}y
      \frac{\partial}{\partial \pi}+\frac{\Pi_{0}}{\mu c\beta^{2}}\frac{\partial}{\partial \tau} \nn \\
 \tilde{X}^{L}_{\pi}&=&\frac{\Pi_{0}}{\mu c}\frac{\partial}{\partial \pi}+
      \frac{mcy}{\Pi_{0}+mc}\frac{1}{\hbar}\Xi  \label{CamposL}\\
 \tilde{X}^{L}_{y}&=&\frac{\Pi_{0}}{\mu c}\frac{\partial}{\partial y}+
       \frac{\pi}{\mu c^{2}\beta^{2}}\frac{\partial}{\partial \tau}-
       \frac{\mu c\pi}{\Pi_{0}+\mu c}\frac{1}{\hbar}\Xi \nn \\
 \tilde{X}^L_\phi&=&\frac{\partial}{\partial\phi}\equiv\Xi\;,\nn
\eea
\noi as well as the right-invariant vector fields, which provide the quantum operators on $U(1)$-complex
functions on the group, once a polarization had been imposed. They are:
\bea
 \tilde{X}^{R}_{\tau}&=&\frac{\partial}{\partial \tau} \nn \\
 \tilde{X}^{R}_{\pi}&=&\frac{\beta}{\mu\omega}\sin\omega \tau\frac{\partial}{\partial y}+
       (\frac{\omega y\pi}{\mu c^{2}\beta}\sin\omega \tau +
        \frac{\Pi_{0}}{\mu c\beta}\cos\omega \tau)\frac{\partial}{\partial \pi}+  \nn \\
      & &\frac{y}{\mu c^{2}\beta}\cos\omega \tau\frac{\partial}{\partial \tau}-
         \frac{1}{(\Pi_{0}+\mu c)\beta}(\Pi_{0}y\cos\omega \tau-\frac{\pi c}{\omega}\sin\omega \tau)
           \frac{1}{\hbar}\Xi          \label{CamposR}\\
 \tilde{X}^{R}_{y}&=&\beta \cos\omega \tau\frac{\partial}{\partial y}+
        (\frac{\omega^{2}y\pi}{c^{2}\beta}\cos\omega \tau-\frac{\Pi_{0}\omega}{c\beta}\sin\omega \tau)
      \frac{\partial}{\partial \pi}-
       \frac{y\omega}{c^{2}\beta}\sin\omega \tau\frac{\partial}{\partial \tau}+ \nn \\
      & &\frac{\mu}{(\Pi_{0}+\mu c)\beta}(\Pi_{0}\omega y\sin\omega \tau+\pi c\cos\omega \tau)\frac{1}{\hbar}\Xi \nn \, .
\eea

The structure of the algebra (\ref{Oscilata}) prevents the existence of a first-order polarization
subalgebra leading to the configuration ``representation'' (there are first-order polarizations
constituted by ladder operators leading to the Fock ``representation'' \cite{oscilata,RMP}). However, it is
possible to look for a second-order polarization subalgebra of the left-enveloping algebra reproducing
the configuration ``representation''. The simplest choice is the algebra generated by:
\be
<\tilde{X}^{HO}_\tau\equiv(\tilde{X}^L_\tau)^2-c^2(\tilde{X}^L_y)^2-\frac{2i\mu c^2}{\hbar}\tilde{X}^L_\tau+
\frac{i\mu c^2\omega}{\hbar}\Xi\,,\,\tilde{X}^L_\pi>\;,
\ee
\noi which must be imposed, along with the $U(1)$-constraint, to complex functions
$\Psi(\phi,y,\pi,\tau)$ on the extended group. The solutions are:
\bea
 \Xi.\Psi=i\Psi&\rightarrow&\Psi=e^{i\phi}\Phi(y,\pi,\tau) \nn \\
 \tilde{X}^{L}_{\pi}.\Psi=0&\rightarrow&\Psi=e^{i\phi} e^{\frac{i}{\hbar}f}\psi(y,\tau) \label{REqXS}  \\
 \tilde{X}^{HO}_{\tau}.\Psi=0&\rightarrow&\frac{1}{\beta^{2}}\frac{\partial^{2}\psi}{\partial \tau^{2}}-
     \frac{2i\mu c^{2}}{\hbar\beta^{2}}\frac{\partial\psi}{\partial \tau}-
     2\omega^{2}y\frac{\partial\psi}{\partial y}-
     c^{2}\beta^{2}\frac{\partial^{2}\psi}{\partial y^{2}}- \nn \\
    & &\frac{\mu^{2}c^{4}}{\hbar^{2}\beta^{2}}\psi+
    \frac{\mu c^{2}\omega}{\hbar}\psi=0 \,, \nn
\eea
\noi where $f$ is the function that appears in (\ref{RGL2}). By restoring the rest-mass
energy\footnote{The actual way of centrally extending a Lie group with trivial cohomology, like the
relativistic symmetry associated with the free particle or the harmonic oscillator, consists
in redefining one particular generator, the energy in this case, with a term proportional to
the central generator.} we get a Klein-Gordon-like equation from the third line in (\ref{REqXS}):
%
\be
 \hat{C}\varphi\equiv-\frac{c^{2}}{\omega^{2}}\Box\varphi=N(N-1)\varphi\, ,
 \label{KGEq}
\ee
\noi where
\be
 \Box\equiv\frac{1}{c^{2}\beta^{2}}\frac{\partial^{2}}{\partial\tau^{2}}-
      \frac{2\omega^{2}y}{c^{2}}\frac{\partial}{\partial y}-
      \beta^{2}\frac{\partial^{2}}{\partial y^{2}} \label{Dalam}
\ee
\noi is the D'Alambertian in an Anti-de Sitter space-time and
$N=\frac{\mu c^2}{\hbar\omega}$; see Ref. \cite{RMP} where the connection to the
motion in a homogeneous space under the group $SO(1,2)$, that is,
the Anti-de Sitter universe, is studied. We use the notation
$\hat{C}\varphi$ to hilight that the l.h.s in (\ref{KGEq})
is the quantum realization of the Casimir operator of the Lie algebra of $SL(2,\mathbb{R})$
\cite{Bacry}.

 The equation (\ref{KGEq}) can be solved in power series. Writing the energy wave functions
in the form
\be
 \varphi_n\equiv e^{-ib_{n}\omega\tau}\beta^{-c_{n}}\HNln\, ,\label{varfi}
\ee
\noi and putting it in equation (\ref{KGEq}), we obtain the relations
\bea
 b_{n}&=&c_{n}  \\
 c_{n}&=&c_{0}+n\equiv N+n \nn \, ,
\eea
\noi as well as the differential equation for the polynomials  $H^N_n$:
\be
 (1+\frac{\zeta^{2}}{N})\frac{d^{2}}{d\zeta^{2}}\HNln-\frac{2}{N}
      (N+n-1)\xi\frac{d}{d\zeta}\HNln+
      \frac{n}{N}(2N+n-1)\HNln=0 \, ,
 \label{HEq}
\ee
\noi where $\zeta\equiv\sqrt{\frac{\mu\omega}{\hbar}}y$.

Equation (\ref{HEq}) defines the so called ``Relativistic Hermite Polynomials'' (RHP) originally
found in Ref. \cite{oscilata} and further developed in Ref. \cite{RMP}. There, we gave the
corresponding Rodrigues' formula:
\be
 \HNln(\zeta)=(-1)^{n}(1+\frac{\zeta^{2}}{N})^{N+n}\frac{d^{n}}{d\zeta^{n}}
          \left[(1+\frac{\zeta^{2}}{N})^{-N}\right]\, .\label{Rodrigues}
\ee

The normalized solutions of eq. (\ref{KGEq}), with respect to the scalar product
\be
<\Psi,\Psi'>=\int\Psi^*(y,\tau)\Psi'(y,\tau)dyd\tau\,, \label{prodesc-xt}
\ee
\noi which is invariant under the group $SL(2,\mathbb{R})$, are given by:
\bea
\Psi^N_n(y,\tau)=C^N_n e^{-i(N+n)\omega\tau}\beta^{-(N+n)}\HNln(\sqrt{\frac{\mu\omega}{\hbar}}y)\,,
\label{normalizadas-xt}
\eea
\noi where
\be
C^N_n=\sqrt{\frac{\omega}{2\pi}}\left(\frac{\mu\omega}{\hbar\pi}\right)^{1/4}
\sqrt{\frac{N^{n-1/2} \Gamma(2N)\Gamma(N)}{n!\Gamma(2N+n)\Gamma(N-\medio)}}\,.
\ee

Creation and annihilation operators for the RHO can be introduced simply  by
$\hat{Z}\equiv  \tilde{X}^{R}_{y} -i\mu\omega \tilde{X}^{R}_{\pi}$ and
$\hat{Z}^\dag\equiv -\tilde{X}^{R}_{y} -i\mu\omega \tilde{X}^{R}_{\pi}$. They turn out to be,
 when acting on the solutions of eq. (\ref{REqXS}) (see \cite{RMP,oscipert}):
\bea
\hat{Z}&=&\sqrt{\frac{\hbar}{2\mu\omega}}e^{i\omega t}\beta\left[\parcial{y}+i\frac{\omega y}{c^2 \beta^2}
\parcial{\tau}+\frac{\mu\omega y}{\hbar\beta^2}\right] \nn \\
\hat{Z}^\dag&=&\sqrt{\frac{\hbar}{2\mu\omega}}e^{-i\omega t}\beta\left[-\parcial{y}+i\frac{\omega y}{c^2 \beta^2}
\parcial{\tau}+\frac{\mu\omega y}{\hbar\beta^2}\right]\,.
\eea

These operators are adjoint to each other with respect to the scalar product (\ref{prodesc-xt}). Their
action on normalized solutions (\ref{normalizadas-xt}) is:
\bea
\hat{Z}\Psi^N_n &=& \sqrt{n\frac{2N+n-1}{2N}}\Psi^N_{n-1} \\
\hat{Z}^\dag\Psi^N_n &=& \sqrt{(n+1)\frac{2N+n}{2N}}\Psi^N_{n+1}\,.
\eea

It should be remarked that the normalized solutions (\ref{normalizadas-xt}) are orthogonal on account 
of the integration in $\tau$, but not in $y$. This causes problems in the time factorization in order
to obtain the minimal (versus manifestly, or time-dependent) realization, in terms of just $y$ 
(see \cite{RMP} for a discussion and \cite{oscipert} for a detailed explanation), and a modification of 
the scalar product and the creation and annihilation operator is needed. In fact, the new scalar product is:
\be
<\Psi,\Psi'>=\int\Psi^*(y)\Psi'(y)\frac{dy}{\beta^2}\,,\label{prodesc-x}
\ee
\noi and the normalized solutions  with respect to this scalar product, with the time dependence
factorized out, are:
\bea
\Psi'^N_n(y)=C'^N_n \beta^{-(N+n)}\HNln(\sqrt{\frac{\mu\omega}{\hbar}}y)\,,
\label{normalizadas-x}
\eea
\noi where
\bea
C'^N_n&=&\sqrt{\frac{N+n}{N-\medio}}C^N_n
= \sqrt{\frac{\omega}{2\pi}}\left(\frac{\mu\omega}{\hbar\pi}\right)^{1/4}
\sqrt{\frac{N^{n-1} (N+n)\Gamma(2N)\Gamma(N+1)}{n!\Gamma(2N+n)\Gamma(N+\medio)}}\nn \\
&=& \frac{1}{2\pi}\sqrt{\omega}\left(\frac{\mu\omega}{\hbar}\right)^{1/4}
2^N N^{n/2}\Gamma(N)\sqrt{\frac{ N+n}{n!\Gamma(2N+n)}}\,.
\eea

The modified creation and annihilation operators, adjoint to each other with respect to the
new scalar product (\ref{prodesc-x}), are obtained \cite{oscipert} through the unitary transformation
$\hat{Z}'=\hat{U}\hat{Z}\hat{U}^{-1}$, $\hat{Z}'^\dag=\hat{U}\hat{Z}^\dag\hat{U}^{-1}$, where
$\hat{U}$ is the unitary operator:
\be
\hat{U}=\sqrt{\frac{\hat{H}}{\hbar\omega(N-\medio)}}e^{\frac{i}{\hbar}\tau(\hat{H}-\mu c^2)}\,,
\ee
\noi where $\hat{H}=i\hbar\parcial{\tau}$ when acting on the manifestly covariant realization, and
an infinite power expansion in $\frac{d\ }{dy}$ and $y$ on the minimal realization (see \cite{oscipert}),
which acquires the simple expression $\hbar\omega(N+n)$ on energy eigenfunctions (\ref{normalizadas-x}). 
The expression of the new ladder operators in the minimal realization, acting on energy eigenfunctions 
(\ref{normalizadas-x}), is \cite{RMP}:
\bea
\hat{Z}'&=&\sqrt{\frac{\hbar}{2\mu\omega}}\sqrt{\frac{N+n-1}{N+n}}\beta\left[\frac{d\ }{dy}+
\frac{\mu\omega y}{\hbar\beta^2}\frac{N+n}{N}\right] \\
\hat{Z}'^\dag&=&\sqrt{\frac{\hbar}{2\mu\omega}}\sqrt{\frac{N+n+1}{N+n}}\beta\left[-\frac{d\ }{dy}
+\frac{\mu\omega y}{\hbar\beta^2}\frac{N+n}{N}\right]\,.
\eea

The RHP have been studied by different authors and related to other already known polynomials,
such as Jacobi \cite{Jacobi} or Gegenbauer \cite{Gegenbauer} polynomials, and the essential of the
latter is here collected since it is relevant for the next section. In fact, in \cite{Gegenbauer}
is proved the actual relation:
\be
H^N_n(u\sqrt{N})=\frac{n!}{N^{\frac{n}{2}}}(1+u^2)^{\frac{n}{2}}C^{\,N}_n(\frac{u}{\sqrt{1+u^2}})\,,
\label{H-C1}
\ee
\noi where $C^N_n(u)$ are the Gegenbauer polynomials \cite{Abramowitz} directly related to the
hypergeometric functions $_2F_1$. For negative index, $N\equiv -q$, we can also write
\be
H^{-q}_n(\sqrt{q}u)\approx C^{\,q-n+\frac{1}{2}}_n(u)\,.\label{H-C2}
\ee
\noi It should be remarked that in Ref. \cite{Gegenbauer} it is
commented that ``$H^N_n(\xi)$ can actually be expressed directly as a (generalized) Gegenbauer
polynomial in the form $C^{-N-n+\frac{1}{2}}(i\xi/\sqrt{N})$. This representation does
not seem to be very useful, however". We shall see in the next section that this connection actually
realizes the analytical prolongation of solutions from the positive to the negative part of the
spectrum of the MPT Hamiltonian.

\section{The Quantum P\"oschl-Teller system}
The commutation relations in (\ref{SO21}) and in (\ref{Oscilata}) are formally analogous
provided that we redefine in (\ref{Oscilata}) the generator $\tilde{X}^R_\tau$ as
$(\tilde{X}^R_\tau)'\equiv\tilde{X}^R_\tau+\mu c^2\tilde{X}^R_\phi$, a redefinition which has been
referred to as the restoring of the rest-mass energy and which, in mathematical terms,
trivializes the central extension of the original $SO(2,1)$ algebra\footnote{The affine form in (\ref{Oscilata}) is needed
to perform the correct non-relativistic limit, which is a group contraction from $SO(2,1)$ to the harmonic oscillator group.}. 
We then aim at finding the quantum theory of the MPT dynamics in the quantum representation
space of this symmetry and resorting to its enveloping algebra in search of the actual
MPT Hamiltonian operator.

Let us proceed in a direct way, once the explicit computations have been developed for the
$SO(2,1)$ group in the example of the relativistic harmonic oscillator. First of all, we restore the
standard notation $t,x,p$ to represent time, coordinate and momentum for the MPT
problem associated with a particle of mass $m$. The essential problem now is to find the
explicit form of the operator $i\hbar\frac{\partial}{\partial t}$, the square of ${\cal E}$,
acting on the wave functions representing the classical Poisson algebra (\ref{SO21}) when
rewritten in terms of the variable $x\equiv \hbox{sinh}^{-1}(\alpha\xi)/\alpha$. To this
end we rewrite (\ref{varfi}) for a negative value $N\equiv -q<0$ of the Bargmann index of
the discrete series of the $SL(2,\mathbb{R})$ representations:
%
\be
 \varphi^q_n\equiv e^{-ic_{n}\omega\tau}(1+\frac{\omega^2}{c^2}y^2)^{-\frac{c_n}{2}}H^{-q}_n\, ,
\ee
\noi or, making explicit the $\hbar$ constant, in terms of $u\equiv\frac{\zeta}{\sqrt{q}}$, $\zeta\equiv\sqrt{\frac{\mu\omega}{\hbar}}y$, and taking into
account that $\frac{\omega^2}{c^2}=\frac{\mu\omega}{\hbar N}$, and $c_n=n-q$,
\be
 \varphi^q_n(\tau,u)=e^{-ic_{n}\omega\tau}\Psi^q_n(u)\equiv  e^{-ic_{n}\omega\tau} (1-u^2)^{\frac{q-n}{2}}H^{-q}_n(\sqrt{q}u)\, .
\label{varfiPT}
\ee

\noi In Table \ref{tabla} the expression of the RHP with different values of
$q$ are shown.
\begin{table}
\begin{tabular}{ccc}
\hline \\
$q=1$ & $q=\frac{3}{2}$ & $q=1.8$  \\
\hline \\
$\ba{lcl}
H^{-1}_0(x)&=&1\\
H^{-1}_1(x)&=&2x \\
H^{-1}_2(x)&=&2(x^2-1)\\
H^{-1}_3(x)&=&0\\
 & \vdots & \\
H^{-1}_n(x)&=&0\,,\,n>3
\ea$
&
$\ba{lcl}
H^{-\frac{3}{2}}_0(x)&=&1\\
H^{-\frac{3}{2}}_1(x)&=&2x \\
H^{-\frac{3}{2}}_2(x)&=&-2+\frac{8}{3}x^2\\
H^{-\frac{3}{2}}_3(x)&=&-4x+\frac{16}{9}x^3\\
H^{-\frac{3}{2}}_4(x)&=&4\\
H^{-\frac{3}{2}}_5(x)&=& -\frac{40}{3}x\\
 & \vdots & \\
H^{-\frac{3}{2}}_{4+k}(x)&\hbox{of} & \hbox{degree $k$}
\ea$
&
$\ba{lcl}
H^{-1.8}_0(x)&=&1\\
H^{-1.8}_1(x)&=&2x \\
H^{-1.8}_2(x)&=& -2 + \frac{26}{9} x^2\\
H^{-1.8}_3(x)&=& \frac{16}{81}x( -27 + 13 x^2 )\\
H^{-1.8}_4(x)&=& \frac{16}{243}( 81 - 54 x^2 + 13 x^4 )\\
H^{-1.8}_5(x)&=& -\frac{32}{2187}x( 405 - 90 x^2 + 13 x^4 )\\
 & \vdots & \\
H^{-\frac{3}{2}}_{n}(x)&\hbox{of} & \hbox{degree $n$}
\ea$ \\
\hline
\end{tabular}
\caption{Expression of RHP with different values of the negative Bargmann index
$-q$:}\label{tabla}
\end{table}

Let us try to derive the Schr\"odinger equation for the MPT potential from the Klein-Gordon equation of
the relativistic harmonic oscillator (with negative Bargmann index). From equation (\ref{KGEq}) we can isolate the second "time"
derivative of $\varphi$:
\be
\frac{\partial^2\varphi}{\partial\tau^2}=
-\omega^2(1-u^2)\left[-2u\frac{\partial\varphi}{\partial
u}+(1-u^2)\frac{\partial^2\varphi}{\partial u^2}+q(q+1)\varphi\right]\,.
\ee

Expressing the $u$-derivative in terms of $x$-derivative, from the relation $u\equiv\hbox{tanh}(\alpha x)$,
\be
\frac{\partial^2}{\partial x^2}=\alpha^2(1-u^2)\left[(1-u^2)\frac{\partial^2}{\partial
u^2}-2u\frac{\partial}{\partial u}\right]\,,
\ee
\noi and defining $D$ through
$q(q+1)\equiv N(N-1)=\frac{2m D}{\alpha^2\hbar^2}=(\frac{2D}{\hbar\Omega})^2$  we obtain:
\be
-\frac{\hbar^2\alpha^2}{2m\omega^2}\frac{\partial^2}{\partial\tau^2}\varphi=
\left[\frac{\hbar^2}{2m}\frac{\partial^2}{\partial
x^2}\varphi+\frac{D}{\hbox{cosh}^2(\alpha x)}\varphi\right]\,.
\ee
\noi This way, denoting $\;t\equiv \frac{2\omega}{\Omega}\tau$,
$\;\hat{E}\equiv i\hbar\frac{\partial}{\partial t}\;$, and defining $\;\varphi\equiv
e^{-\frac{i}{\hbar}\sqrt{D}\sqrt{-E}t}\chi$, $\;(-E\equiv\epsilon)\;,$ we arrive at
the time-independent Schr\"odinger equation for a particle of mass $m$ in a
MPT potential with depth $D$ and width $1/\alpha$:
\be
\frac{\hbar^2}{2m}\frac{d^2\chi}{dx^2}+\left(E+\frac{D}{\hbox{cosh}^2(\alpha
x)}\right)\chi=0\,.
\ee

The solutions to this equation were given in terms of Gegenbauer polynomials \cite{Arias}:
\be
\chi^q_n(u)\approx(1-u^2)^{\frac{q-n}{2}}C^{\, q-n+\frac{1}{2}}_n(u)\,,
\ee
\noi which can now be compared  with the time-independent part, $\Psi^q_n(u)$, of the functions of 
the relativistic harmonic oscillator (\ref{varfiPT}) (through the relation (\ref{H-C2})).

The scalar product for the minimal (time-independent) realization can be directly derived or obtained from
that of the RHO, changing $\frac{dy}{\beta^2}\rightarrow \frac{du}{1-u^2}$:
\be
<\Psi,\Psi'>=\int_{-1}^{1}\Psi^*(u)\Psi'(u)\frac{du}{1-u^2}\,. \label{productoescalar}
\ee

The ladder operators for this system can also be obtained from the ones for the RHO given in the previous
section, simply by changing $N\rightarrow -q$ and performing the appropriate change of variables.  
On eigenfunctions they have the expression:
\bea
\hat{Z}'&=&
    \frac{1}{\sqrt{2q}}\sqrt{\frac{q-n+1}{q-n}}\sqrt{1-u^2}\left[\frac{d\ }{du}+\frac{u}{1-u^2}(q-n)\right]\nn\\
\hat{Z}'^\dag&=&
    \frac{1}{\sqrt{2q}}\sqrt{\frac{q-n-1}{q-n}}\sqrt{1-u^2}\left[-\frac{d\ 
    }{du}+\frac{u}{1-u^2}(q-n)\right]\,,\label{ladder}
\eea
\noi and the action of these operators on normalized eigenstates have the simple form:
\bea
\hat{Z}'\chi^q_n=\sqrt{n\frac{2q-n+1}{2q}}\chi^q_{n-1}\\
\hat{Z}'^\dag\chi^q_n=\sqrt{(n+1)\frac{2q-n}{2q}}\chi^q_{n+1}\,.
\eea

These operators and their action coincide, up to a constant factor, with 
the ones given in \cite{Lemus}.

From the action of the ladder operators, we conclude that the representation space has dimension $2q+1$ (for $q$ integer or half-integer), since
$\hat{Z}\chi^q_0=0$ and $\hat{Z}^\dag\chi^q_{2q}=0$. Unlike the RHO, the MPT has only
a finite number of (bounded) states.

The spectrum of the MPT Hamiltonian can also be derived from that of the RHO:
\be
E_n=-\frac{\hbar^2\alpha^2}{2m}(q-n)^2=-\frac{\hbar^2\Omega^2}{4D}(q-n)^2=-\frac{D}{q(q+1)}(q-n)^2\, ,\qquad n=0,1,\ldots, 2q\,.
\label{spectrum}
\ee

Let us look in detail at the obtained representation. We shall first consider the case of integer $q$.
From equation (\ref{Rodrigues}) with $N=-q$, we observe that $H^{-q}_n=0$ for
$n>2q$ (see Table \ref{tabla}). Therefore there are just $2q+1$ states, in agreement with the previous
statement that the representation is finite-dimensional.
Then, if $q$ is an integer, all eigenvalues except one are doubly degenerated, the
minimum being $E_0= -\frac{\hbar^2\alpha^2}{2m}q^2=-\frac{q}{q+1}D=E_{2q}$, and the maximum being $E_q=0$. However, this
degeneracy is only apparent, since the complete wave function for the states with the same energy,
$\Psi^q_n$ and $\Psi^q_{2q-n}\,,n=0,\ldots,q-1$, are identical. Furthermore, if we consider the normalization
of the states with the scalar product (\ref{productoescalar}), it turns out that the state $\Psi^q_q$, the one with zero energy,
is not normalizable. This means that the physical Hilbert space is spanned by $\Psi^q_n\,,n=0,\ldots,q-1$,
since the other states, $\Psi^q_n\,,n=q+1,\ldots,2q$, are copies of them (we can also think of it
as if they were not reachable by the action of creation operators, since the state $\Psi^q_q$ is
out of the Hilbert space).

If $q$ is half-integer, from equation (\ref{Rodrigues}) with $N=-q$ (see Table \ref{tabla}), we deduce that there are an infinite
number of states. Their behavior is as follows: for $n=0,\ldots,2q$, $H^q_n$ is a polynomial of degree $n$,
as should be, but $H^q_{2q+1}$ is a polynomial of degree zero, and then $H^q_{(2q+1)+k}\,,k=1,2,\ldots$
is a polynomial of degree $k$. However, by the action of the ladder operators only the first $2q+1$
states are reachable, and the representation is finite-dimensional.
Even more, the states $\Psi^q_n$ and $\Psi^q_{2q-n}\,,n=0,\ldots,q-\medio$ are not identical and therefore
 there is a double degeneracy for all the states. However, if we take into account the normalizability
 with respect to the scalar product (\ref{productoescalar}), it turns out that the physical Hilbert space
 is spanned by $\Psi^q_n\,,n=0,\ldots,q-1/2$, the rest of states being not normalizable.

In summary, for the finite-dimensional (non-unitary) representations of $SL(2,\mathbb{R})$,
from the $2q+1$ states of the representation, $\Psi^q_n,\,,n=0,\ldots,2q$, only $[q]+1$ are
normalizable, $\Psi^q_n,\,,n=0,\ldots,[q]$, where $[q]$ stands for the smaller, closest integer to $q$, 
and these span the physical Hilbert space. These states are also orthogonal with respect to the scalar product (\ref{productoescalar}), and the orthonormal
basis is:
\be
\tilde{\Psi}^q_n(u)=N^q_n\, \Psi^q_n(u)\,,\,N^q_n=\frac{ 2^{-q} }{\Gamma(q+1)} \sqrt{\frac{\left( q - n \right) \,
          \Gamma(2\,q - n+1)}{n!}}\,,\quad n=0,\ldots,[q]\,.
\ee

If we express the solutions in terms of Gegenbauer polynomials, the results are similar; the
only difference is that for the non-normalizable states they are not defined. The reason is that the
proportionality constant in (\ref{H-C2}) diverges for these cases.

These features are very different from that of $SU(2)$ representations, which are also finite-dimensional,
but unitary and, therefore, for $j$ integer or half-integer, all $2j+1$ states are orthogonal and
normalizable. This clearly implies that we cannot use $SU(2)$ as the symmetry group for bounded states.
 Furthermore, the use of $SU(2)$ leads to
inconsistencies, since it predicts a double degeneracy in the eigenstates, something that it is
forbidden in one dimension.
Despite of this, it has been widely used in the literature, see for instance \cite{Iachello,Arias}.

This results can be extended to the Morse Potential \cite{Rosen-Morse,Iachello,Arias2}.
As in the present case there is a finite number of bounded states, which are associated with a
finite-dimensional, non-unitary representation of $SL(2,\mathbb{R})$ (although in the literature they
have also been associated with $SU(2)$).

An important fact of having finite-dimensional representations of $SL(2,\mathbb R)$ instead of $SU(2)$ is 
that, going to the universal covering group of $SL(2,\mathbb R)$, all real values of $q$ are allowed.
In this case (see Table \ref{tabla}) $H^{-q}_n$ is a polynomial of degree
$n$ for all $n\in\mathbb N$, but 
taking into account the normalizability with respect to the scalar product
(\ref{productoescalar}), only the first $[q]+1$ states are normalizable,
from $n=0,1,\ldots,[q]$, and these spand the physical Hilbert space\footnote{This is in agreement with
the WKB counting of bounded states for a general potential \cite{Galindo}, applied to the P\"oschl-Teller 
potential, which turns out to be $N\approx \medio+\sqrt{q(q+1)}$, and this equals $q+1$ for large $q$.}. 
Since $SU(2)$ is already simply-connected, no real values other than integer or half-integer are allowed 
for the index $j$ labelling its representations.
 This has relevant consequences from the physical point of view. Since
 $q(q+1)=\frac{2m D}{\alpha^2\hbar^2}=(\frac{2D}{\hbar\Omega})^2$, the restriction of $q$ to integer
 and half-integer values (as happens for $SU(2)$ representations) leads
 to a formal quantization of the potential parameter $D/\alpha^2$ (or rather $\frac{2D}{\hbar\Omega}$), 
whereas this does not happen for finite-dimensional $SL(2,\mathbb R)$ representations,
 where all real values of $q$ are allowed.
 
Probably, the most important reason to support the idea of describing the bounded states of the MPT system
by $SL(2,\mathbb R)$ instead of $SU(2)$ is the harmonic limit, which consists in taking $D\rightarrow\infty,\,\alpha\rightarrow 0$ such
that $\alpha^2D$ is kept constant. Both the positive discrete series
and the finite-dimensional representations of $SL(2,\mathbb R)$ contract, under the limit 
$N\rightarrow\infty$ and $q\rightarrow\infty$, respectively, to the harmonic oscillator. In fact, from
eq. (\ref{Rodrigues}) it can be directly checked that 
${\rm lim}_{N\rightarrow\pm\infty}H^N_n(x)=H_n(x)$.  For the case of the finite-dimensional representations, 
the harmonic limit of the energies requires a previous redefinition, in such a way that 
${\rm lim}_{D\rightarrow \infty} (E_n+D) = \hbar\Omega(\medio+n)$, that is, the spectrum of the 
harmonic oscillator with frequency $\Omega=\omega(D)$ is recovered. Even the ladder operators (\ref{ladder}) goes to the
ladder operators of the harmonic oscillator (with frequency $\Omega$) in the harmonic limit (see \cite{Lemus}). However,  
contracting the $SU(2)$ representations to that of the harmonic oscillator would require a negative
spin index.
 
As a last general comment, we should say that a more complete study of the P\"oschl-Teller dynamics resorting
to the GAQ of the $SL(2,\mathbb C)$ group would be in order. In that case, the different parts of the
spectrum would be more properly related to different (real) subgroups.

\end{document}